\documentclass[12pt]{article}
\usepackage{amsmath,amssymb,natbib}
\newcommand{\kms}{\ensuremath{\,\mbox{km}\,\mbox{s}^{-1}}}
\newcommand{\hii}{H\,{\sc ii}}
\newcommand{\aap}{A\& A}

\newcommand{\apj}{ApJ}
\newcommand{\apjl}{ApJL}
\newcommand{\apjs}{ApJS}

\newcommand{\mnras}{MNRAS}

\newcommand{\araa}{ARA\&A}
\newcommand{\aj}{AJ}
\newcommand{\aaps}{A\& AS}
\begin{document}
\title{The Radio-FIR correlation: Is MHD Turbulence the Cause?}
\author{Brent A. Groves$^{1}$, Jungyeon Cho$^{2}$,
Michael Dopita$^{1}$ \& Alex Lazarian$^{2}$
}
\date{}
\maketitle
{\center
$^{1}$ {Research School of Astronomy \& Astrophysics, Australian National University,
Cotter Road, Weston Creek, ACT 2611 \\
bgroves@mso.anu.edu.au\\}
$^{2}$ Univ. of Wisconsin, Madison WI53706, USA
}
%
%\altaffiltext{1}{Research School of Astronomy \& Astrophysics,
%Australian National University,
%Cotter Road, Weston Creek, ACT 2611
%Australia}
%\altaffiltext{2}{Univ. of Wisconsin, Madison WI53706, USA}
%\altaffiltext{3}{bgroves@mso.anu.edu.au}
%
%
\begin{abstract}
The  radio  -  far  infrared   correlation  is  one  of  the  tightest
correlations found  in astronomy. Many  of the models  explaining this
correlation  rely  on the  association  of  of  global magnetic  field
strength  with gas density.  In this  letter we  put forward  that the
physical  reason for  this association  lies within  the  processes of
magnetohydrodynamic turbulence.
\end{abstract}
\bigskip
\section{Introduction}

One  of  the more  extraordinary  correlations  in  astronomy is  that
between  the far  infrared (FIR)  and radio  continuum of  galaxies. A
correlation  between the  IR emission  and  radio was  first noted  in
Seyferts by  \citet{vdKruit71}, it  was not until  the results  of the
IRAS mission  were analysed that universality of  this correlation was
discovered    \citep{DickeySal84,deJKW85,HelouSRR85}.    This   linear
correlation spans  $\sim 5$  orders of magnitude  with less  than 50\%
dispersion  \citep{YunRedCon01,Wunderlich87},  making  it one  of  the
tightest correlations  known in astronomy.  It appears  to be followed
by all galaxies with ongoing star-formation and without a dominant AGN
\citep{Niklas97}. The correlation not  only holds on global scales but
is also found to hold within  the disks of galaxies, down to scales on
the  order   of  a  few   100  pc  \citep{BeckGol88}.  What   is  most
extraordinary  about this  relationship is  that it  couples  a purely
thermal process  in the  IR (the re-emission  of UV radiation  by dust
grains)  with   a  non-thermal  process  at   radio  wavelengths  (the
synchrotron  radiation  of  relativistic  electrons).   Typically,  at
1.4~GHz, the  non-thermal emission dominates  by at least an  order of
magnitude         over          the         free-free         emission
\citep{Condon92,Niklas97,Dopita02},  and  hence  the  contribution  of
free-free emission to the correlation is minimal.

The radio --  FIR correlation is not a  simple mass-scaling (richness)
effect \citep{Wunderlich88,XuLVW94},  and hence must  be explained by some  form of
direct  coupling between  the  IR emitting  dust  and the  synchrotron
emission of cosmic  rays.  Star formation is generally  accepted to be
responsible for the correlation \citep{Wunderlich88}. The  basic scenario is as follows: As
the massive, hot stars are born in star-forming regions and live their
short lives,  the UV radiation  they emit heats the  surrounding dust,
which then re-emits in the IR. This radiation also provides the energy
for  the thermal  (free-free) radio  emission.  When  these  hot stars
reach the end of their brief existence, the resulting supernova shocks
create energetic cosmic rays which  are believed to be responsible for
the non-thermal radio emission.  However,  as it stands this model has
too many steps and too many parameters to explain the tightness of the
correlation  (see  the illustration  by  \citet{Ekers91}).  Thus  more
complex  models have  been put  forward,  although there  is still  no
consensus on the cause of the radio -- FIR correlation, or an adequate
explanation for  its tightness.   In several of  these models  a large
part  of  the  explanation  of  the correlation  was  related  to  the
association of global magnetic field strength with gas density through
an  energy equipartition  scheme.  In  this  letter we  put forward  a
physical  mechanism  for this  assumption  adopted  in these  previous
models.  We  show here  that the cause  for this relationship  may lie
within the processes of magnetohydrodynamic (MHD) turbulence.

In section 2 we discuss the previous theoretical models that have been
previously  proposed to  explain  the correlation.   In  section 3  we
discuss  MHD turbulence  and how  it naturally  provides  the required
relationship  between  magnetic field  strength  and  gas density.  In
section 3 we  discuss how this relation provides  a possible basis for
the correlation,  and how it fits  into the previous  models, with the
concluding remarks in section 4.

\section{Previous Theoretical Models}

As  mentioned in  the introduction,  recent star-formation  provides a
basis for most theoretical models of the FIR-radio correlation. One of
the earliest  theories, the `optically-thick'  or `calorimeter' theory
\citep{Volk89,VolkXu94,LisenVX96}, assumed  three things.  First, that
all the FUV\footnote{FUV radiation is that which lies between $\approx
5$ eV and $\approx 13.6$  eV} radiation from massive stars is absorbed
by the dust grains within a galaxy.  Second, that the energetic cosmic
rays produced by the supernova  explosions of these stars lose most of
their energy within the galaxy  due to synchrotron and inverse Compton
process. As  both these  processes are proportional  to the  number of
massive   stars,   these   calorimetric   assumptions  lead   to   the
correlation. Finally, the tightness  of the correlation is provided by
the  third assumption  that  the energy  density  of the  interstellar
radiation field, $U_{rad}$,  is in a constant ratio  with the magnetic
field energy density,  $U_B$.  In other models which  do use the first
assumption  (`optically thin'  models), including  ours, there  is the
generally implicit  assumption that there is a  direct linear relation
between  the gas  density and  dust  density, an  assumption which  is
nonetheless supported by observations \citep{XuHel96}.

An  alternative theory  put  forward by  \citet{HelBic93} assumes  the
opposite extreme, an `optically-thin'  model, in which the cosmic rays
and UV  photons both  have high escape  probabilities. To  provide the
correlation  in these  optically and  CR thin  galaxies they  have two
assumptions.  Firstly that the  UV or  `dust-heating' photons  and the
radio emitting cosmic rays are  created in constant proportion to each
other,  which  is  again  related  to star  formation.  Secondly,  the
tightness of the  correlation is provided by a  local coupling between
the magnetic field strength and the gas density.

A   challenge    to   both   these   theories   is    the   model   by
\citet{NikBec97}. In  this work they argue  that observations indicate
that within most galaxies, CR electrons lose very little energy before
they escape.  These same galaxies  are optically thick to  UV photons,
thus  both  the  calorimetric   and  optically--thin  models  are  not
supported  by  the observations.  In  the  Niklas  \& Beck  model,  the
controlling factor they put forward  for the correlation is the volume
density of the gas. They assume that both the star formation rate (and
thus dust  heating) and the magnetic field  strength (which determines
the synchrotron emission) depend upon the gas volume density and hence
the correlation.

Of  course  there  have  been   other  approaches,  such  as  that  of
\citet{Bettens93}  who  looked  at  cosmic  ray  driven  chemistry  in
molecular clouds  to explain the  correlation. However, in  their case
they did not take account of the strength of the interstellar magnetic
field, an important factor in the correlation.

A  different  way  to  look   at  the  correlation  was  suggested  by
\citet{HoeBX98} when examining M31. They decomposed both the radio and
the FIR into two parts. The  radio was decomposed into a thermal radio
component and  a non-thermal component,  while the FIR  was decomposed
into  a warm  component, associated  with  \hii\ regions,  and a  cool
component, associated  with the  diffuse (cirrus) clouds.   They found
both  a correlation  between  the thermal  radio  and warm  FIR and  a
correlation  between the  non-thermal  radio and  the  cool FIR  (like
\citet{Xu94}).  The first correlation is easily understandable as both
components are associated with  \hii\ regions.  The second correlation
is not so easy to understand, as  the cool FIR component of M31 is not
believed to  be predominantly heated  by massive stars. The  local low
mass  stars  provide  most of  the  heating,  yet  these are  not  the
progenitors of supernovae which provide the CR electrons necessary for
the  non--thermal radio.   Thus in  M31 at  least, the  correlation of
non-thermal radio --  cool FIR must be due to  factors other than star
formation. What \citet{HoeBX98} suggest is that the correlation is due
to the close coupling of  the magnetic field strength and gas density.
Assuming that the gas density is proportional to the dust density then
the synchrotron emission (via  the magnetic field) is directly related
to the FIR emission (due to the local dust density).

It is clear that most of  these theories require some form of coupling
between the  magnetic field and  gas density.  It is  usually asserted
that  `equipartition' provides  the  coupling, yet  the  way in  which
equipartition can arise  is left open. In the  next section we explore
whether MHD turbulence can provide the necessary coupling mechanism.

\section{Turbulent Coupling}

\subsection{MHD Turbulence}

Turbulent motions are observed  in most astrophysical fluids and since
magnetic   fields   are    undoubtedly   present   in   such   fluids,
magnetohydrodynamic (MHD) turbulence is an important field of study in
astronomy (see \citet{ChoLV02} for a review of MHD turbulence).

In  most astrophysical  plasmas, the  magnetic Reynolds  number ($R_m=
L\delta V/\eta$, where $L$ is the size of the system, $\delta V$ is 
the the r.m.s. velocity, and $\eta$ is the magnetic diffusion) easily
exceeds $10^{10}$ and the usual  expectation is that magnetic field is
frozen into the  gas in such systems.  The  velocity field advects and
stretches magnetic field lines  and magnetic field exerts pressure and
tension   forces   on   velocity   fields.   With   these   additional
considerations, MHD  turbulence is  generally different from  the pure
hydrodynamic case.

On the  scales over which  the radio --  FIR correlation holds,  it is
reasonable to assume that the over-all magnetic field is weak. In this
weak/zero mean  field regime,  there are two  main mechanisms  for the
generation     of     magnetic     field:    the     dynamo     effect
\citep{Parker79,Moff78}  and  field  line stretching  \citep{Batch50}.
The dynamo effect  can amplify the mean magnetic  field ($B_0$), while
field line  stretching is responsible  for amplification of  the local
magnetic field.

Using  MHD turbulent  models, \citet{ChoVish00}  have shown  that even
without  the dynamo  effect,  field line  stretching  can amplify  the
magnetic fields up to the level of energy equipartition.  According to
this model, the  rate of field line stretching at  the scale of energy
injection  or the  largest  energy containing  eddies,  $L$, is  $\sim
\delta V/L$, while the rate  of magnetic back-reaction at the scale is
$\sim (\delta  B/\sqrt{4 \pi \bar{\rho}})/L$, where $\delta  V$ is the
r.m.s.velocity, $\delta B$ is  the r.m.s. magnetic field strength, and
$\bar{\rho}$ is the average density.

Therefore,  when  $\delta B/\sqrt{4  \pi  \bar{\rho}}  \ll \delta  V$,
stretching  is more  effective  than back-reaction,  resulting in  the
growth of the r.m.s. field strength. The stretching is balanced by the
back-reaction only when the local Alfv\'en velocity ($\delta B/\sqrt{4
\pi \bar{\rho}}$) is comparable with the local fluid velocity ($\delta
V$).   In other  words, magnetic  field cannot  further grow  by field
stretching  when  the   energy  equipartition  condition,  $\bar{\rho}
(\delta V)^2 \sim  (\delta B)^2/(4 \pi)$, has been  reached.  That is,
when MHD turbulence reaches  the stationary state, the magnetic energy
density   matches   the   energy   density   of   the   gas,   $\delta
B/\sqrt{4\pi\bar{\rho}}\sim  \delta V$.  In  the weak  field case  the
turbulent  local  magnetic  field  ($\delta  B$) is  larger  than  the
external, mean magnetic field ($B_0$).

In principle,  the dynamo can  amplify the large-scale  magnetic field
($B_0$).  In mean field dynamo theory (see Moffatt 1978; Parker 1979),
turbulent  motions at  small scales  are biased  to create  a non-zero
electromotive force  along the  direction of the  large-scale magnetic
field.   This  effect (the  `$\alpha$-effect')  works  to amplify  and
maintain  large-scale magnetic  fields.   Whether or  not this  effect
actually  works  depends  on  the  structure of  the  MHD  turbulence,
especially  on  the  mobility   of  the  field  lines.   For  example,
\citet{VainCat92} have argued that when equipartition between magnetic
and  kinetic energy  densities occurs  at  any scale  larger than  the
dissipation scale,  the mobility of  the field lines and  the $\alpha$
effect will be greatly reduced.  However the nature and degree of this
suppression is  a controversial issue  (see \citet{GruDia94, CatHug96,
BlackFie00,   VishCho01};   see   \citet{VishLC02}  for   a   review).
Nevertheless, when the mean field grows to the value similar to energy
equipartition ($B_0/\sqrt{4 \pi  \bar{\rho}} \sim \delta V$), mobility
of magnetic  field lines is  greatly reduced and therefore  the dynamo
can      no      longer      operate.       Numerical      simulations
(e.g.~\citet{ChoVish00b}) show that,  when $B_0/\sqrt{4 \pi \rho} \sim
\delta V$, there also exists (almost exact) energy equipartion between
random magnetic and turbulent  kinetic energy.  From this observation,
we can assume that  $B_0/\sqrt{4 \pi \bar{\rho}} \sim \delta B/\sqrt{4
\pi \bar{\rho}}\sim  \delta V$.  Therefore, when the  dynamo operates,
the local magnetic field  ($B\sim \sqrt{B_0^2+(\delta B)^2}$), as well
as the fluctuating one ($\delta B$), stays at the equipartition value:
$B/\sqrt{4 \pi \bar{\rho}} \sim \sqrt{2}~ \delta V$.

Thus, regardless  of the strength of the  external (galactic) magnetic
field, the  local magnetic field  will stay in equipartition  with the
gas;

\begin{equation}\label{equipartition}
\delta V \sim B/\sqrt{4\pi\overline\rho}.
\end{equation}
This provides  the coupling between  the gas density and  the magnetic
field  which is  a necessary,  but not  sufficient, condition  for the
operation of the radio -- FIR correlation.

\label{section3.1}

\subsection{Magnetic field and Gas Density Coupling}

When  the velocity  dispersion of  quiescent  gas within  the disk  of
galaxies  is  measured it  is  found  to  be remarkably  constant.  HI
observations of the LMC found  the velocity dispersion lies within 6.8
and 7.7 \kms across the galaxy \citep{Kim99}.  Similar observations of
the  nearly  face-on   galaxies  NGC  628  and  NGC   3938  also  show
approximately  constant velocity  dispersion  across the  face of  the
galaxies  in   both  CO   (NGC  628:  6\kms,   NGC  3938:   8.5  \kms,
\citet{CombesBec97}) and HI  (NGC 628: $\sim 9\kms$ \citep{Shostak84},
NGC  3938: $\sim 10\kms$  \citep{Kruit82}). This  trend of  a velocity
dispersion  of $\sim 10  \kms$ continues  beyond these  three objects,
with the  dispersion being appreciably uniform  between many galaxies,
including       both      spirals       and       irregulars      (see
e.g.~\citet{SellBal99,Kamph93}).    Though  the  reason   behind  this
unusual constancy is not yet  understood, we can still apply this fact
to the situation of energy equipartition.

Equation \ref{equipartition} with  a constant velocity dispersion then
implies that:

\begin{equation}\label{Brho}
      B \propto \sqrt{\overline\rho}.
\end{equation}

This    relation   is    consistent   with    other    numerical   MHD
simulations. These  simulations of turbulent interstellar  gas show $B
\propto {\overline\rho}^m$ with  $m$ found to be $m  \sim $0.4--0.6 by
several independent  models \citep{PadNor99,KimBalMac01,OstSG01}.  The
relation  between $B$  and  ${\overline\rho}$ shows  that  there is  a
tendency  toward energy equipartition  even within  individual eddies,
which  may   represent  individual  clouds.   Note   that  the  energy
equipartition  described  in  \S\ref{section3.1}  is  for  the  entire
system, which  may represent  units as large  as an entire  galaxy. Of
course, for  this to represent  an entire galaxy  we must rely  on the
implicit assumption that the magnetic field within the halo is closely
coupled to that  within the disc.

A  similar correlation  between $\overline  B$ and  $\overline\rho$ is
also seen observationally within galaxies, with \citet{Berk97} showing
$\overline B  \propto {\overline\rho_{g}}^m$ in M31 and  the Milky Way
with  $m  \sim  $0.3--0.7.   Similarly, there  lies  an  observational
correlation  between  $\overline  B$  and  $\overline\rho$  on  global
galactic  scales, with \citet{NikBec97}  using 43  galaxies to  find a
relation  with  a slope  $m=0.48\pm0.05$.   Note  that  in both  cases
equipartition between cosmic rays  and magnetic field energy densities
was assumed.

Thus the relation  between $B$ and $\rho$ is  seen in observations and
is     also     inferred      in     our     numerical     simulations
\citep{ChoVish00,ChoVish00b},   with  MHD  turbulence   providing  the
mechanism for this coupling.

\section{The Connection to the Radio -- FIR Correlation}

As discussed in \S 2, most  models that provide an explanation for the
radio -- FIR  correlation assume a coupling of  the magnetic field and
gas density of the form of eqn. \ref{Brho}. The physical mechanism put
forward  in the previous  section justifies  this assumption  and thus
provides further credence to these models.

What the  relation between  $B$ and $\rho$  also provides is  a direct
coupling between two parameters which  determine the amount of flux in
the FIR and radio and hence relates directly to the correlation. If we
ignore the  effect of  the thermal radio  emission - which  is usually
minimal on galactic scales\citep{Condon92,Niklas97,Dopita02}, then the
remaining parameters of the correlation  are the number density of the
cosmic rays and the dust-heating UV radiation field.

The  density  of  cosmic   rays  is  presumably  determined  by  their
production  via  Fermi  acceleration  processes  occurring  in  strong
shocks, mostly generated by stellar winds and by supernova explosions,
their radiative lifetime,  and the volume of the  galaxy in which they
can   move.    If,   in   a   disk   galaxy,   $   \stackrel{.}{\Sigma
}_{\mathrm{SF}}$ is the birthrate of young stars per unit area of the
disk,  and  the characteristic  scale-height  of  the  cosmic rays  is
$z_{\mathrm{  cr}}$ ,  determined  by the  large-scale magnetic  field
configuration, then the number density of comic rays with be given by

\begin{equation}
n_{\mathrm{cr}}=\phi \stackrel{.}{\Sigma
}_{\mathrm{SF}}/z_{\mathrm{cr}}.
\end{equation} 
where  $\phi  $  is  a   constant  determined  by  the  ratio  of  the
characteristic    acceleration    efficiency    and   the    radiative
lifetime. These radiative losses will mostly occur in highly localized
regions  corresponding to the  denser neutral  clouds where  the local
magnetic field has been enhanced by
 MHD turbulence. The local synchrotron emissivity at frequency $\nu $
is given by
\begin{equation} 
j_{\nu }=f(a)kB^{\left( a+1\right) /2}\nu ^{-\left(
a-1\right) /2} 
\end{equation} 
where  the   number  density   of  the  relativistic   electrons  with
relativistic  $\gamma  $  has   a  power  law  distribution  $N(\gamma
)^{-a}=k\gamma ^{-a}$ , and $B$ is the local magnetic field. The total
density    of     the    relativistic    electrons     is    therefore
$n_{\mathrm{cr}}=k\gamma  _{\min  }^{1-a}/\left(  a-1\right)  $  where
$a\sim  2.4$  which  corresponds  to  a frequency  spectral  index  of
synchrotron emission $\alpha =-0.7$. Such a value for the index is not
unreasonable \citep{Stevens02},  though the value could  be flatter or
steeper than this. Therefore, the non-thermal emissivity scales as the
product $n_{\mathrm{cr}}B^{1.7},$ approximately, or:
\begin{equation} 
j_{\nu }=\phi B^{1.7}\stackrel{.}{\Sigma
}_{\mathrm{SF}}/z_{\mathrm{cr}} \label{Radio} 
\end{equation}

The  local infrared emissivity  scales as  the local  radiation field,
which can be approximated as:
\[
n_{\mathrm{UV}}=\psi (\stackrel{.}{\Sigma}_{\mathrm{SF}}
+ \zeta {\Sigma}_{\mathrm{*}})/z_{\mathrm{*}},
\]
where $\psi$ and $\zeta$  are scaling factors, $z_{\mathrm{*}}$ is the
characteristic    scale    height    of    the   young    stars    and
${\Sigma}_{\mathrm{*}}$  is  the  surface  density of  the  cool,  old
stellar  population. Note  that the  scale height  of the  old stellar
population is larger than that of the young population but this factor
is taken account within $\zeta$.  The local emissivity also depends on
the local gas density and the radiation field-weighted mean opacity of
the dust  grains, $\left\langle \kappa  _{\mathrm{UV}}\right\rangle $,
which though dependent upon  the grain properties and radiation field,
will  vary only  to  a small  extent.   Thus, assuming  that the  dust
emission is optically thin, we have
\begin{equation} 
j_{\mathrm{FIR}}=n\psi \left\langle \kappa
_{\mathrm{UV}}\right\rangle (\stackrel{.}{\Sigma }_{\mathrm{SF}}+
\zeta {\Sigma}_{\mathrm{*}})/z_{\mathrm{*}} .
\label{IR} 
\end{equation} 
Therefore, with our MHD turbulence, which provides the
scaling relationship $ n\propto B^{2},$ locally, we would expect that
\begin{equation} 
\frac{j_{\nu }}{j_{\mathrm{FIR}}}=\frac{\phi
z_{\mathrm{*}}\stackrel{.}{\Sigma }_{\mathrm{SF}}}{\psi \left\langle
\kappa _{\mathrm{UV}}\right\rangle z_{\mathrm{cr}}
(\stackrel{.}{\Sigma }_{\mathrm{SF}}+
\zeta {\Sigma}_{\mathrm{*}})}n^{0.15}. 
\label{relation}
\end{equation} 
This  provides the  required radio-FIR  correlation to  the  degree to
which the product of the physical parameters on the right hand side of
this equation remain constant.  Of  course this is only valid when the
same   volume  element   produces   both  the   FIR  and   synchrotron
emission. There  are possibly  other volume elements  (such as  in the
halo)  which  might  only  produce  radio emission.  This  means  that
equation \ref{relation} may not  be relevant for the global FIR--radio
correlation,  only  giving  a  quantitative  analysis  for  the  local
correlation.   Additionally, variations  in these  physical parameters
with different  galaxy types may also explain  the non-linearities and
outliers in  the correlation, such  as starbursts \citep{YunRedCon01}.
Further  work  needs  to  be  done  on  the  relation  to  the  global
correlation  as well  as the  scatter  in the  parameters in  equation
\ref{relation} and others, like the local synchrotron index, to
determine  their  effects upon  the  correlation  and whether  further
constraining  mechanisms are needed  to explain  the tightness  of the
correlation.

Thus,  though  we  have  explained  one  part  of  the  radio  --  FIR
correlation,  more  understanding  is   needed  of  the  cosmic  rays,
radiation field  heating the dust,  and the processes  discussed above
before the exact  reason for the correlation and  its tightness can be
fully comprehended.

\section{Conclusion}

Most models  which try to  explain the remarkable  correlation between
radio and  FIR emission  rely upon the  association of  magnetic field
strength  and gas  density.   MHD simulations  show  that through  the
process  of equipartition, the  magnetic field  and gas  densities are
coupled, with
\begin{displaymath}
B \propto \sqrt{\bar{\rho_g}},
\end{displaymath}
a relationship that  is also indicated by observations  within our own
and other galaxies.  This relationship  provides a basis for the radio
-- FIR  correlation by directly  connecting one  of the  parameters of
synchrotron  emission with  a parameter  of FIR  emission.   While not
explaining the  relationship fully,  it does put  us a step  closer to
fully understanding this remarkable correlation.

\section*{Acknowledgments}  The authors  would like  to thank  the  referees for
several insightful  comments and suggestions which  improved the paper
greatly.
BG would like to acknowledge the assistance of the
Alex Rodgers  Traveling Scholarship and Edward  Courbold Research Fund
for this  collaboration. MD wishes  to acknowledge the support  of the
Australian  National University  and the  Australian  Research Council
(ARC) under  his ARC Australian Federation Fellowship,  and also under
ARC Discovery project DP0208445.

\end{document}